# The Star Formation Histories of Disk and E/S0 Galaxies from Resolved Stars

Knut A.G. Olsen
National Optical Astronomy Observatory
kolsen@noao.edu
Phone: (520)-318-8555

Co-authors:
Aaron J. Romanowsky (UCO/Lick)
Abhijit Saha (NOAO)
Evan Skillman (University of Minnesota)
Ben F. Williams (University of Washington)
Rosemary F.G. Wyse (Johns Hopkins University)

Submitted to GAN Panel

**Summary**
The resolved stellar populations of local galaxies, from which it is possible to derive complete star formation and chemical enrichment histories, provide an important way to study galaxy formation and evolution that is complementary to lookback time studies. We propose to use photometry of resolved stars to measure the star formation histories in a statistical sample of galaxy disks and E/S0 galaxies near their effective radii. These measurements would yield strong evidence to support critical questions regarding the formation of galactic disks and spheroids. The main technological limitation is spatial resolution for photometry in heavily crowded fields, for which we need improvement by a factor of ~10 over what is possible today with filled aperture telescopes.

**Background**
One of the great unknowns in modern astronomy is how galaxies form and evolve within the context of the large-scale structure of the expanding universe. In recent years, a profitable approach to the problem has been to observe ensembles of galaxies forming and evolving as a function of redshift, as done through e.g. the Hubble Deep Fields (Williams et al. 1996, 2000), the Ultra Deep Field (Beckwith et al. 2006), and numerous ground-based imaging and spectroscopic surveys. There remain large uncertainties in our knowledge, however, despite the progress of these lookback time studies. One issue is that high redshift galaxies are essentially viewed as snapshots in time, which complicates efforts to piece together the evolution of the population as a whole. A second issue is that high redshift galaxies cannot be resolved into individual stars, such that interpretations of the observations depend on expectations formed from observations of local galaxies. Our knowledge of the stellar populations of local galaxies is, in turn, also quite incomplete. In particular, there is currently very little overlap between the kinds of galactic environments that we can study locally and those that are seen at high redshift.

The study of galaxy formation and evolution rests, ultimately, on our application of the theory of stellar evolution to the light that we observe from distant galaxies. The most basic analysis that we can make, therefore, of a galaxy's evolutionary history is to measure its resolved stellar populations and use our understanding of stellar structure and evolution to derive its stellar age and metallicity distributions (Tolstoy & Saha 1996, Dolphin 1997, Harris & Zaritsky 2001). With this approach, we can reconstruct the entire star formation histories of single galaxies, rather than limited snapshots in time, as well as identify potential inadequacies in the theoretical models themselves. Because the most uncertain phases of stellar evolution are typically those that are the shortest lived and produce the brightest stars, this resolved star analysis is critical for interpreting the light received from distant unresolved galaxies, where the brightest stars have disproportionate influence. The main drawback is that the approach is currently only feasible in the nearest galaxies and, for high surface brightness components of galaxies such as bulges and inner disks, severely limited even at the cosmologically negligible distance of M31. As a consequence, we do not have anything like a statistical or representative sample of galaxies available to us for this otherwise powerful technique. In our view, developing technologies that would greatly extend our ability use resolved stars to dissect the star formation histories of galaxies should be a high priority for the coming decade. In particular, we set as a goal the measurement of star formation histories of a sample of galaxy disks and bulges and of massive elliptical galaxies. Reaching this goal would bridge the current gap between studies of local and high redshift galaxies, placing both on firmer footing.

**How and when did galaxies form stable disks? What are the star formation histories of elliptical and S0 galaxies?**

The formation of galaxy disks is a critical topic in galaxy formation. The observations that most of the stellar mass at z~0 resides in galaxies with masses of at least $5\times10^{10}$ $M_{sun}$ (Kauffmann et al. 2003), that massive "red sequence" galaxies cannot have formed in gas-rich mergers (Faber et al. 2007), and that most currently active star-forming galaxies are disk galaxies suggests that the majority of stars once formed in disks. Disks are also the components most easily destroyed in the hierarchical formation of galaxies, such that their survival is sensitive to the details of galaxy formation physics (Hopkins et al. 2009a). Several simulations of disk galaxy formation have shown that the physics of stellar feedback and the numerical resolution of the simulations play critical roles in determining the structures, dynamics, and star formation and chemical enrichment histories of the simulated disks (Abadi et al. 2003, Robertson et al. 2004, Governato et al. 2007). While there are sufficient integrated light data for comparing the broad properties of simulated disk structures and dynamics to observations (e.g. Giovanelli et al. 1997) precise star formation and chemical enrichment histories for a sample of galaxy disks simply do not exist; these would be extremely valuable. In addition, with knowledge of the star formation histories of a statistical sample of disks, we would be able to observe directly trends with galaxy mass and density of the environment, which are predicted in hierarchical assembly (Springel, Frenk, & White 2006) and are key to connecting detailed knowledge of local galaxies to the galaxy population as a whole.

Another crucial issue for galaxy formation is when and under what circumstances did elliptical and S0 galaxies form the bulk of their stars. The consensus paradigm is well represented by the analysis and review of Faber et al. (2007), who outline the many possible processes involved in the transition of E/S0 galaxies from actively star forming states to their current inactive ones. Of particular importance are gas-poor mergers for massive ellipticals, gas-rich mergers for low-mass ellipticals, and a variety of star formation quenching mechanisms for all ellipticals and S0's. As described by Hopkins et al. (2009b), E/S0 galaxies at z=0 are each multicomponent mixes of stellar populations formed by the sum of these processes. Mapping the complete star formation and chemical enrichment histories in a number of E/S0 galaxies would provide a way to identify the bursts of star formation expected to be seen in their dissipational components. As seen for the case of dwarf galaxies, high quality data of nearby dwarfs have been crucial for understanding their formation (Orban et al. 2008), and falsifying what had been an established paradigm (Dekel & Silk 1986).

**Method and recent results**

The basic approach to the problem of extracting the complete star formation and chemical enrichment history from resolved star photometry is to identify the linear combination of stellar evolutionary model isochrones that provides the best statistical description of observed color-magnitude diagrams (e.g. Dolphin 2002). The quality of the results depends on getting observations deep enough to sample a wide range of phases of stellar evolution (including the red giant and AGB phases, the horizontal branch stars, and, if possible, the main sequence turnoff of the oldest stars) and wide enough to contain large samples of stars. Spectroscopy of individual stars is also valuable for providing complementary constraints on the observed chemical abundance distributions.

Current facilities are able to approach the problems of disk and E/S0 formation through resolved stars only in the most nearby ($D<\sim 4$ Mpc) galaxies, and then only by either restricting observations to very low surface brightness regions of the galaxies, or by using only the brightest phases of stellar evolution, such as the RGB and AGB phases (observations of pure galaxy halos *are* possible to larger distances, but these do not address the questions posed here). The reason for the constraints is the severe crowding in the high surface brightness regions of nearby massive galaxies. As argued by Olsen, Blum, & Rigaut (2003), crowding stems from the fact that when performing stellar photometry, there is no way to remove the star of interest and measure the contribution to its luminosity of the background underneath. Instead, the *average* background must be estimated from regions adjacent to the star, which will typically differ from the *true* background by the characteristic size of random luminosity fluctuations within a single resolution element of the telescope. This line of reasoning leads to a direct relationship between the luminosity of and distance to the star, the surface brightness and luminosity function of the stellar population in which the star is embedded, the resolution of the telescope, and the level of acceptable photometric error. In the next paragraphs, we describe the empirical limits of current facilities; these are in excellent agreement with analytical predictions from the simple consideration of crowding described above.

The need for spatial resolution has made the *Hubble Space Telescope* one of the pre-eminent facilities for measuring star formation histories in nearby galaxies. The ACS Nearby Galaxy Treasury Survey (ANGST; Dalcanton et al. 2009) contains the largest, most uniform set of *HST* images of nearby galaxies; it provides multicolor photometry of the stellar populations in a nearly volume-limited sample of galaxies within ~4 Mpc. Williams et al. (2009) used ANGST photometry of a field in the low surface brightness outer disk of M81 ($D$~3.8 Mpc), including stars fainter than the horizontal branch, to measure its star formation history. They found that 60% of the stellar mass formed at least ~8 Gyr ago, with rapid metal enrichment to [M/H] >~ –1; this implies that the bulk of M81's disk was in place by z~1. This important conclusion depends, however, on the assumption that the outer disk field, located at a radius of 5 scale lengths and which contains only a tiny fraction of the disk's stellar mass, is representative of the disk as a whole. As seen in M31 (Ibata et al. 2007), the outer disks of many spiral galaxies may be products of satellite accretion rather than probes of disk formation.

Rejkuba et al. (2005) obtained a deep HST ACS image of a field in the halo of NGC 5128, ~7 $r_e$ from the center of the nearest massive E/S0 galaxy. The color-magnitude diagram, which reaches below the horizontal branch, shows that the field contains stars with a wide range in metallicity, with an average [M/H] of –0.64 dex and a dispersion of ~0.5 dex; Rejkuba et al. determined the average age of the stars to be ~8±3 Gyr. WFPC2 observations of a field ~1.3 $r_e$ from the center, although of much shallower depth and lower quality due to increased crowding, suggest a somewhat higher average metallicity [M/H] of −0.46 dex (Harris & Harris 2002). The average stellar age and possible population gradient argue that NGC 5128 formed the bulk of its stars *in situ* by z~1, after which star formation was quenched. However, this conclusion is again based primarily on data obtained in the outskirts of the galaxy.

The deep HST ACS observations of the spheroid of M31 (Brown et al. 2008), 10–35 kpc from the galaxy center, demonstrate the excellent accuracy with which the star formation and chemical enrichment history can be measured from data reaching below the oldest main sequence turnoffs. Brown et al. found that all of the fields contained significant numbers of ~9–11 Gyr-old stars, with the majority of stars younger than the globular cluster 47 Tuc. One of the fields, located in M31's giant stellar stream (Ferguson et al. 2002), contained the signature of that stream as an overabundance of ~8 Gyr-old stars in the derived star formation history. The Brown et al. results provide convincing evidence for the formation of M31's outer spheroid by satellite accretion.

8-10-m ground-based telescopes equipped with adaptive optics systems are now poised to achieve still deeper photometry in crowded fields, as they are delivering images with resolution equal to or better than HST. Davidge et al. (2005) and Olsen et al. (2006) observed several fields in the M31 bulge and inner disk with Gemini North/NIRI+Altair, with the delivered PSFs having FWHM~0.7 arcsec in H and K. The star formation histories derived from the numerous RGB and AGB stars in the fields showed that both the bulge and inner disk populations in M31 are dominated by old (~8 Gyr) solar-metallicity stars, providing further evidence that massive disks were in place by z~1; this result depends, however, on our uncertain understanding of the AGB phase of stellar evolution, and needs to be verified with deeper photometry.

**The need for higher spatial resolution and sensitivity**
Clearly, more powerful facilities than currently available are needed to be able to measure the star formation and chemical enrichment histories in high surface brightness disks and E/S0 galaxies, mainly to lower the fundamental limit set by crowding. The only paths forward are to build larger facilities with near diffraction-limited imaging capabilities or to improve the image quality of existing large telescopes at the shortest possible wavelengths. A viable solution is to depend on the continued growth of ground-based adaptive optics technology; another would be to build a new large optical space telescope. In Figure 1, we show the limits imposed by crowding, following the analytical approach of Olsen, Blum, & Rigaut (2003), for galaxies at a variety of distances observed with 10, 25, and 30-m telescopes operating at wavelengths corresponding to the V, I, and K bands, all assumed to be diffraction-limited. The surface brightnesses for this calculation were set to $\Sigma_K$=19, $\Sigma_I$=20.3, and $\Sigma_V$=22 mags arcsec$^{-2}$, typical of the disks of spiral galaxies and of ~3 $R_e$ in E galaxies. Setting as our goal to be able to resolve horizontal branch stars, as done by the ANGST survey, in several high surface brightness disks and E/S0 galaxies, we identify two technological ground-based advances that would expand our current sample to include galaxies as far away as the Sculptor group:
1) Developing optical (V-band) adaptive optics for existing ~10-m telescopes
2) Building a 25-30m GSMT equipped with near-infrared (JHK) AO

A more ambitious goal of building a 25-30m GSMT operating near the diffraction limit in the *I* band would revolutionize our ability to measure star formation histories in external galaxies; such a telescope would allow us to resolve the old main sequence in Sculptor group galaxy disks, the subgiant branch in Cen A, and horizontal branch stars about as far away as the Virgo cluster. A large optical space-based telescope would also be ideal for this science, as long as it had an aperture of at least 10-m.

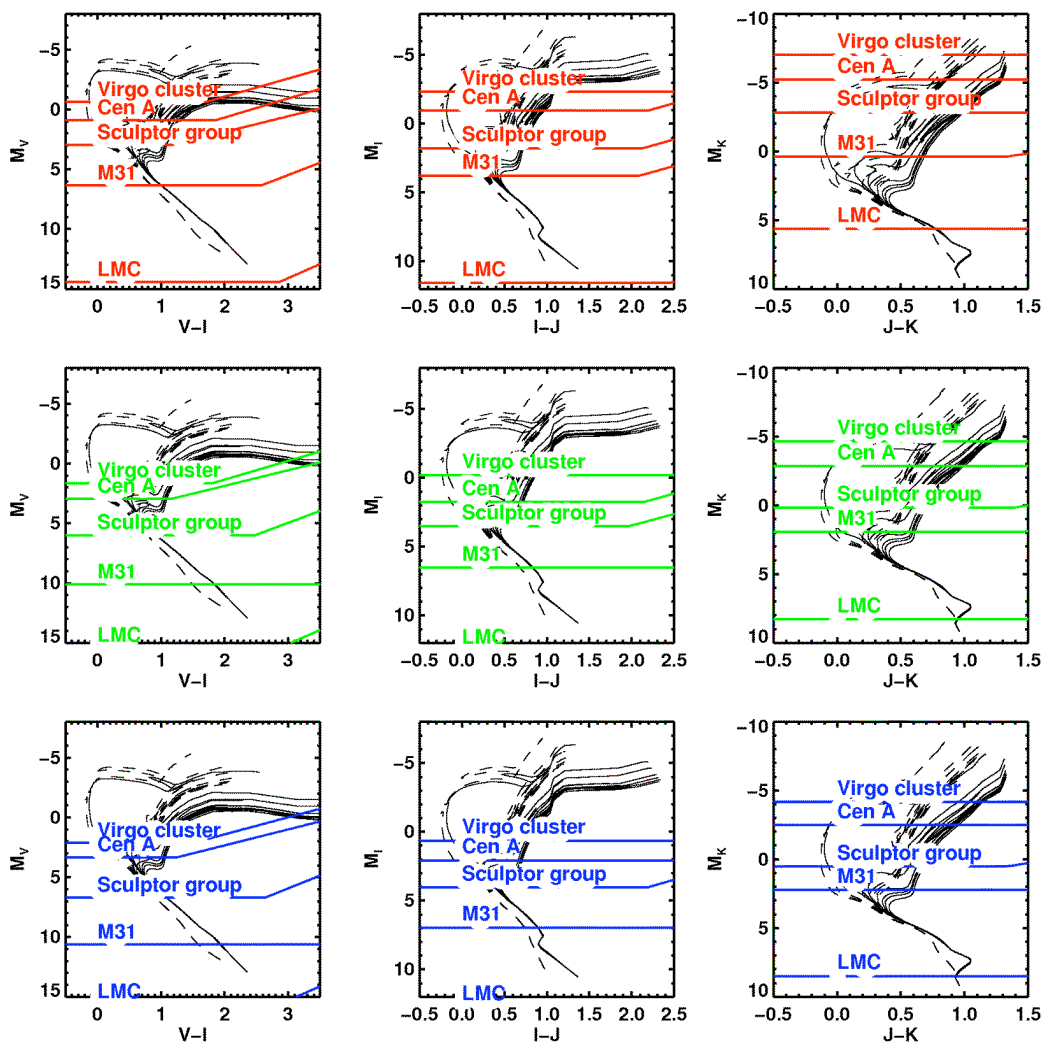

Fig. 1 Limits imposed by crowding (horizontal lines) for galaxies at a variety of distances, observed with 10-m (top row), 25-m (middle row), and 30-m (bottom row) telescopes, and operating at wavelengths corresponding to the V (left column), I (middle column), and K (right column) bands, all assumed to be diffraction-limited. The limits are compared to isochrones ranging in age from 10 Myr to 14 Gyr, and for [M/H]=0 and -1. The surface brightnesses for this calculation were set to $\Sigma_K$=19, $\Sigma_I$=20.3, and $\Sigma_V$=22 mags arcsec$^{-2}$. A 10-m telescope diffraction-limited at *V* and a 25-m to 30-m diffraction-limited at *JHK* would both reach the goals described in this paper. A 25-m to 30-m telescope diffraction-limited at *I* would transform the science, as it would allow us to study horizontal branch stars in disks and E galaxies at ~3R$_e$ in the Virgo cluster.

To demonstrate just how powerful these capabilities would be, Figure 2 shows an observation of the center of the nearby dwarf elliptical galaxy M32 obtained with Gemini North/Hokupa'a+QUIRC as well as simulations of the performance of JWST and a 30-m near-IR GSMT. The JWST color-magnitude diagram goes almost no deeper than the Hokupa'a+QUIRC CMD, revealing only the brightest evolved stars, while the GSMT observation approaches the old main sequence turnoff. The gain in the recovered star formation history provided by the GSMT in this case is clear; whereas JWST provides the correct qualitative features, GSMT measures the star formation history with superb precision. The next decade could truly be a transformative one for our view of the local Universe.

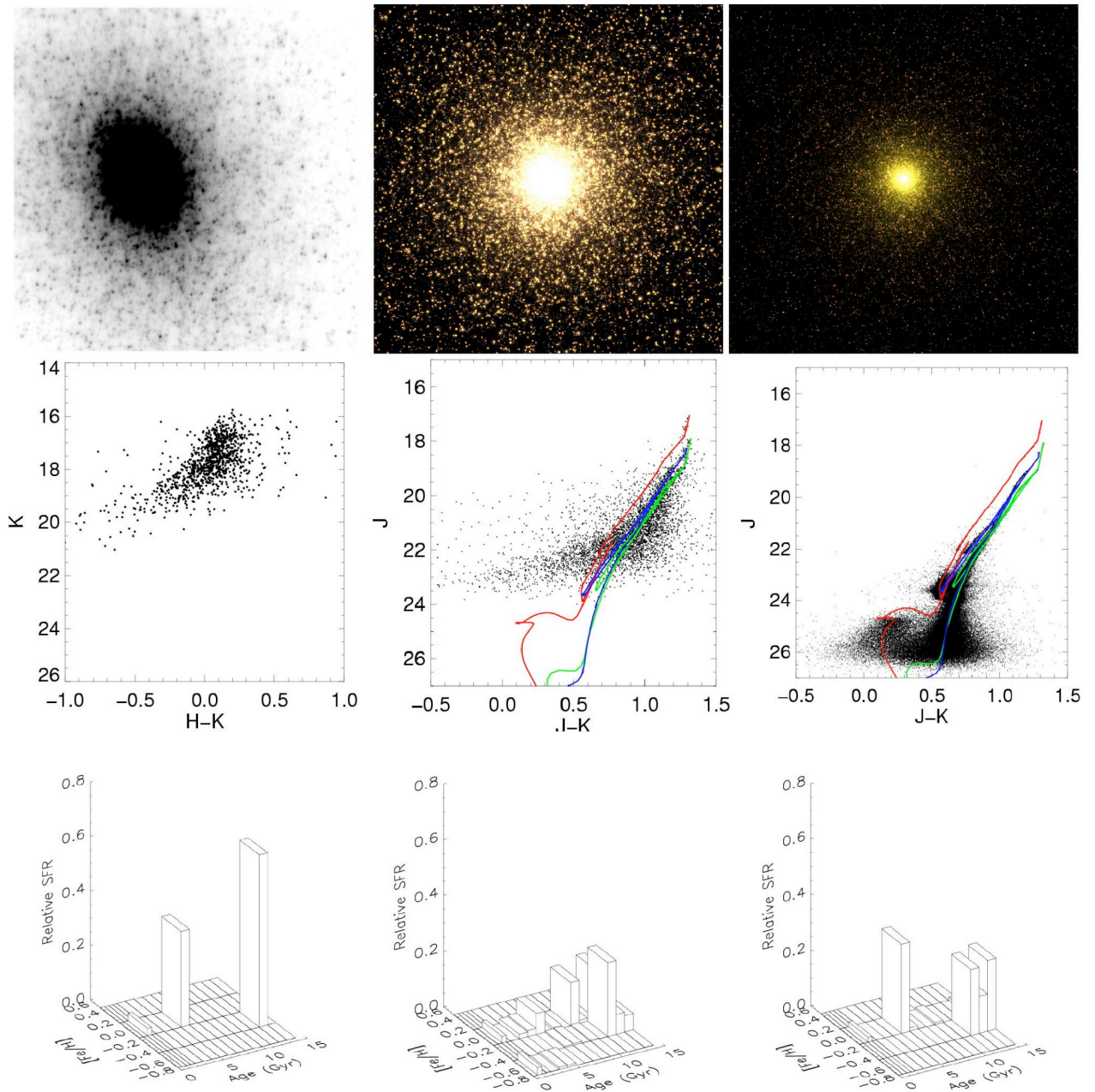

Fig. 2. Stellar populations at the center of M32. Top row, left to right: images of the central 30″ of M32 as observed with Gemini N+Hokupa'a (Davidge et al. 2000) and as simulated with JWST (middle) and a 30-m GSMT (right). Middle row, left to right: Color-magnitude diagrams of M32 corresponding to the images in the top row. Bottom row: The population box used to create the JWST and GSMT simulations is shown at left, while the recovered JWST and GSMT population mixes are shown in the middle and right panels.